\documentclass[%
 reprint,
 amsmath,amssymb,
 aps,
prb,
]{revtex4-2}

\usepackage{graphicx}% Include figure files
\usepackage{dcolumn}% Align table columns on decimal point
\usepackage{bm}% bold math
\usepackage{lmodern}
\usepackage{multirow,booktabs}
\usepackage{bm}
\usepackage{times}
\usepackage{mathtools}
\usepackage{bbold}
\usepackage[utf8]{inputenc}
\usepackage[qm]{qcircuit}
\usepackage{physics}
\usepackage[table]{xcolor}

\usepackage{lastpage}
\usepackage{enumitem}
\usepackage{fancyhdr}
\usepackage{mathrsfs}
\usepackage{wrapfig}
\usepackage{setspace}
\usepackage{calc}
\usepackage{cancel}
\usepackage[retainorgcmds]{IEEEtrantools}
\usepackage{amsmath}

\usepackage{empheq}
\usepackage{framed}
\usepackage[most]{tcolorbox}
\usepackage{xcolor}
\usepackage{algorithm}
\usepackage{algpseudocode}

\newcommand{\cur}[1]{\mathcal{#1}}

\newcommand{\expt}[1]{\left\langle #1 \right\rangle}

\begin{document}

\preprint{APS/123-QED}

\title{Automated Characterization of a Double Quantum Dot using the Hubbard Model}%

\author{Will Wang}
 \email{qinghe1230@ucla.edu}
\affiliation{%
 Physics and Astronomy Department, University of California, Los Angeles
}%

\author{John Dean Rooney}
\affiliation{%
 Physics and Astronomy Department, University of California, Los Angeles
}%

\author{Hongwen Jiang}
\affiliation{%
 Physics and Astronomy Department, University of California, Los Angeles
}%

\date{\today}% It is always \today, today,
             %  but any date may be explicitly specified

\begin{abstract}
Semiconductor quantum dots are favorable candidates for quantum information processing due to their long coherence time and potential scalability. However, the calibration and characterization of interconnected quantum dot arrays have proven to be challenging tasks. One method to characterize the configuration of such an array involves using the Hubbard model \cite{Sarma2011, Wang2011, Yang2011}. In this paper, we present an automated characterization algorithm that efficiently extracts the Hubbard model parameters, including tunnel coupling and capacitive coupling energy, from experimental stability diagrams. Leveraging the dual annealing optimizer, we determine the set of Hubbard parameters that best characterize the experimental data. We compare our method with an alternate, well-established measure of the tunnel coupling and find good agreement within the investigated regime \cite{Wei2013}. Our extracted tunnel couplings range from 69 to 517 $\mu$eV, and we discuss the limiting factors of our method.
\end{abstract}

%\keywords{Suggested keywords}%Use showkeys class option if keyword
                              %display desired
\maketitle

%\tableofcontents

\section{Introduction}

Semiconductor quantum dots are among the most promising hardware platforms for quantum information processing. In the recent years, both single and two-qubit operations using electron spins have been carried out with fidelity above $99\%$ \cite{Mills2022, Yang2017}. Electron spins in silicon also have long coherence time, making them robust to decoherence effects observed in other solid state systems \cite{Yoneda2018}. Furthermore, fabrication of these devices can be easily integrated with the classical semiconductor industry, making semiconductor qubit a true candidate for scalable quantum computing.

However, calibrating a quantum dot array to a desired working point has proven to be nontrivial, due to the large parameter space and each quantum dot having unique parameters compared to its neighbors. This makes the characterization of these devices especially important. To achieve single and two-qubit operations in a quantum dot array, it is imperative to know the tunnel coupling between two neighboring quantum dots. Furthermore, recent works had demonstrated the possibility of simulating the Hubbard model with a quantum dot array \cite{Hensgens2017}. To achieve a direct mapping between the a quantum dot array and the Hubbard model, it is also essential to know the values of the Hubbard model parameters at a given voltage configuration.

One of the primary distinctions of the Hubbard model from the capacitive model is the account of tunnel coupling between neighboring sites. Conventionally, the tunnel coupling energy is extracted from the expectation value of the double dot charge polarization averaged over a Maxwell-Boltzmann distribution \cite{DiCarlo2004, Wei2013}. This measurement, which we will henceforth label as the DiCarlo method, is performed by measuring the electron temperature and the polarization line width at the desired anti-crossing in stability diagrams. To obtain an accurate measurement, additional experimental setups are required for measuring the electron temperature; furthermore, adjustments to the lock-in voltage are needed to bring out the anti-crossing polarization line on the stability diagram. Additionally, at the limit of extremely large tunnel coupling, this method becomes difficult to perform due to extensive broadening of the polarization line.

Another parameter in the Hubbard model that is crucial for studying the coupling between neighboring quantum dots is the capacitive coupling energy. It has been shown that this coupling energy can be directly obtained by measuring the length of the anti-crossing while converting from voltage to energy space with the appropriate lever arms \cite{Neyens2019}. However, this method is limited to the low tunnel coupling region where tunnel effect remains small. The addition of tunnel coupling distorts and increases the spacing of the anti-crossings, making the isolation of the capacitive coupling energy difficult through this simple measurement \cite{Wang2011}.

Inspired by the works of Das Sharma et al. \cite{Sarma2011, Wang2011, Yang2011}, we developed a novel method that automatically extracts the Hubbard model parameters directly from experimental stability diagrams. As discussed in \cite{Wang2011}, information regarding the Hubbard model are contained in the geometry of the stability diagrams. Building on the work of \cite{Wang2011} in this paper, we study the possibility of extracting the Hubbard parameters directly from the stability diagrams with depth and compare our results, specifically the tunnel coupling measurement, to the DiCarlo method. We have found agreement with the DiCarlo method; however, we will also discuss the limitation of our method. Further, we demonstrate our method's ability to extract the capacitive coupling energy in the large tunnel coupling limit, which has yet to be done before from solely the geometry of the stability diagram.

We acknowledge that there have been previous work done on characterizing/calibrating the state of a quantum dot array including the measurement of either tunnel coupling and capacitive coupling energy. For instance, in \cite{VanDiepen}, the authors proposed an automated procedure for tuning the tunnel coupling. While in \cite{Nurizzo2022}, the authors discussed tuning the tunnel coupling of a quantum dot array. However, both papers are different from our approach, where we obtained the tunnel coupling directly from the curvatures of the Hubbard model. Additionally, although \cite{VanDiepen} mentions fitting the anti-crossing, their fitting is fundamentally different from our approach which involves the Hubbard model. In another work \cite{Liu2022}, an automated tuning protocol was proposed. While this paper differentiates tunnel coupling regions with the curvatures of the anti-crossings, its measurements are purely qualitative and are unable to identify the exact tunnel coupling value.

\section{Background}

\subsection{Experimental Setup}
\begin{figure*}
    \centering
    \includegraphics[width=1.0\textwidth]{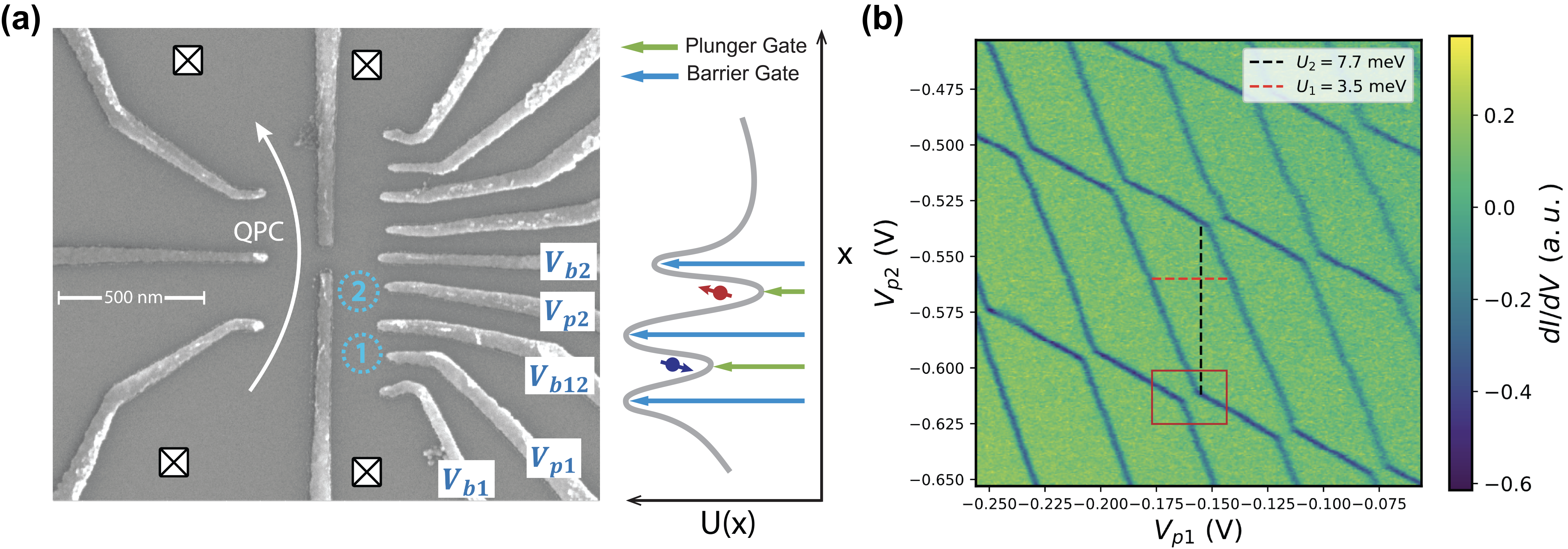}
    \caption{\label{fig1}{\textbf{(a)}: \textbf{Left}: SEM diagram of a the quantum dot array that was used to perform the measurements in this paper. The two blue dotted circles are approximately the positions of the quantum dots, and the QPC is labeled on the left side of the device. \textbf{Right}: A two-site Hubbard model illustration: the grey line depicts the potential landscape of the two dots. The blue and green arrow underneath represents the effects of plunger and barrier gates. In the Hubbard model, plunger gates would be directly related to the chemical potential of each dots; and the barrier gate is used to tune the coupling energy as well as the tunnel coupling. Conceptually, the charging energies of the dots correspond to the size of the dots. \textbf{(b)}: A large stability diagram in the low tunneling regime containing multiple anti-crossings. The dark pixels indicate a rapid change in the current of the QPC. We pick the anti-crossing in the red square as the one to be investigated (this choice is arbitrary). Note that the color bar will be omitted for the rest of the paper. The charging energies for each dot can be directly measured as the distance between horizontal (red, QD1) and vertical (black, QD2) transition lines. The cross-capacitances can be measured from the slopes of the transition lines.}}
\end{figure*}

For this paper, we used a device (Fig. \ref{fig1}) with two quantum dots and an adjacent quantum point contact (QPC), which are defined by gate electrodes fabricated on top of a Si/SiGe heterostructure. The two-dimensional electron gas was formed by a positive voltage applied to a global top gate located above the gate electrodes and insulated by a 100 nm layer of aluminum oxide. Note that the measurements in this paper were performed only with the two lower dots of the device, while the upper portion was left completely open. The two lower dots are coupled by barrier gate $V_{B12}$. The QPC gates are tuned such that the current flowing through the left channel is sensitive to electron transport in the two dots.

The chemical potential of dot 1(2) was primarily controlled through plunger 1(2) above the dot. By tuning the plunger voltages, electrons can hop on and off the dots from the source and drain reservoirs and in between each dot. Electron tunneling was detected as a change in the transconductance signal measured by a lock-in connected to the QPC, while the lock-in’s excitation voltage was applied to both plungers. As shown in Fig. 1b, electron transport in a double quantum dot can be well visualized with stability diagrams created from pairwise scanning of neighboring plungers.

To determine each plunger’s lever arm, we followed the procedure in \cite{Wei2013}. We measured the full width at half maximum (FWHM) of the polarization line widths for both double dots as a function of fridge temperature $T_{f}$. For high $T_{f}$, the polarization line’s FWHM becomes proportional to $T_{f}$, allowing lever arms to be extracted from the proportionality constant. The extracted lever arms values are $0.1\pm0.02\: \textrm{eV}/\textrm{V}$ for both plunger gates to their respective dots.

\subsection{The Hubbard Model}

Conventionally, it is common to model the dynamics of a quantum dot array with the capacitive model \cite{VanDerWiel}. While this model captures the effects of Coulomb blockade and the capacitive coupling between coupled quantum dots, it fails to account for the quantum fluctuation. In order to incorporate the quantum effects in our models, we need to use the extended Hubbard model, a model often used to describe interacting particles on a lattice. To apply the extended Hubbard model to a quantum dot array, we use the following Hamiltonian \cite{Wang2011}:
\begin{multline}
    \hat{\cur{H}} = -\sum_i \mu_i\hat{n}_i - \sum_{\expt{i,j},\sigma} t_{ij}\left(\hat{c}^\dag_{i\sigma}\hat{c}_{j\sigma} +  \hat{c}^\dag_{j\sigma}\hat{c}_{i\sigma} \right) \\+ \sum_i \frac{U_i}{2}\hat{n}_i(\hat{n}_i-1) + \sum_{ i\not=j}U_{ij}\hat{n}_i\hat{n}_j
\end{multline}

\noindent where $\hat{c}_{i\sigma}$ and $\hat{c}^{\dagger}_{i\sigma}$ are the fermionic creation and annihilation operators at site $i$ with spin $\sigma$ (either $\uparrow$ or $\downarrow$); the number operator is defined as $\hat{n}_{i\sigma}=\hat{c}^{\dagger}_{i\sigma}\hat{c}_{i\sigma}$; $\epsilon$ is the single-particle energy offset; $t_{ij}$ is the tunnel coupling between dots $i$ and $j$; $U_{i}$ is the charging energy on dot $i$; $U_{ij}$ is the capacitive coupling energy between dot $i$ and $j$; and finally $\expt{i,j}$ denotes neighboring sites $i$ and $j$. Putting things in a broader context - the first term in the Hamiltonian refers to the single-particle energy offsets; the second term is the hopping term that describes the tunneling effects between neighboring dots; the third term accounts for the on-site Coulomb interaction; and the last term is the inter-site Coulomb interaction. Note that the Hubbard model at the limit of zero tunnel coupling is completely equivalent to the capacitive model \cite{Yang2011}. Also note that in the generic Hubbard model, there are terms that describe the effects of spin exchange, co-tunneling, etc; however, we will ignore these extra terms as their effects are insignificant in our experiment.

Generating stability diagrams from the Hubbard model is discussed in \ref{SimulateSD} and \ref{ConvertVE}. In this work, we also developed a few techniques to speed up the simulation, as discussed in \ref{Speedup}.

\section{Method}
In this section, we will discuss the characterization protocols we have developed as well as the details of the optimization, including a discussion on optimizers.
\subsection{Characterization Procedures}
Here, we provide a brief overview of the entire characterization procedure. We start with generating a large stability diagram in the low-tunneling regime and choose a specific anti-crossing as shown in Figure \ref{fig1}b.

We then measure the vertical and horizontal spacings between the transition lines to obtain the charging energies at low tunnel coupling. These measurements are taken in the voltage space, and are converted to energy units with the appropriate lever arms. For our devices, the lever arms are approximately $0.1 \textrm{ eV}/\textrm{V} = 100 \textrm{ meV}/\textrm{V}$. The measured charging energies for the dot 1 and 2 are $3.5$ meV and $7.7$ meV.

In these devices, cross-capacitances contribute to the slopes of the transition lines. Therefore, we can also measure the cross-capacitances of the two plunger gates from the slopes of the transition lines in the figure above \cite{Volk2019}. Here, we find the cross-capacitances of dot 1 and 2 with respect to plunger gate $V_{p2}$ and $V_{p1}$ to be 0.56 and 0.42.

Since the cross-capacitance and the charging energy can be easily obtained from features in the stability diagram, such as the slope and distance between charging lines, we will keep them as fixed variables in the optimization. A two-dot Hubbard model consists of 8 free parameters, including 2 charging energies, 2 cross-capacitances, 2 voltage offsets, the tunnel coupling, and the capacitive coupling energy. For the rest of the characterization, we will determine the four remaining parameters, that is the capacitive and tunnel coupling, and the two voltage offsets.

\begin{figure*}
    \centering 
    \includegraphics[width=1\textwidth]{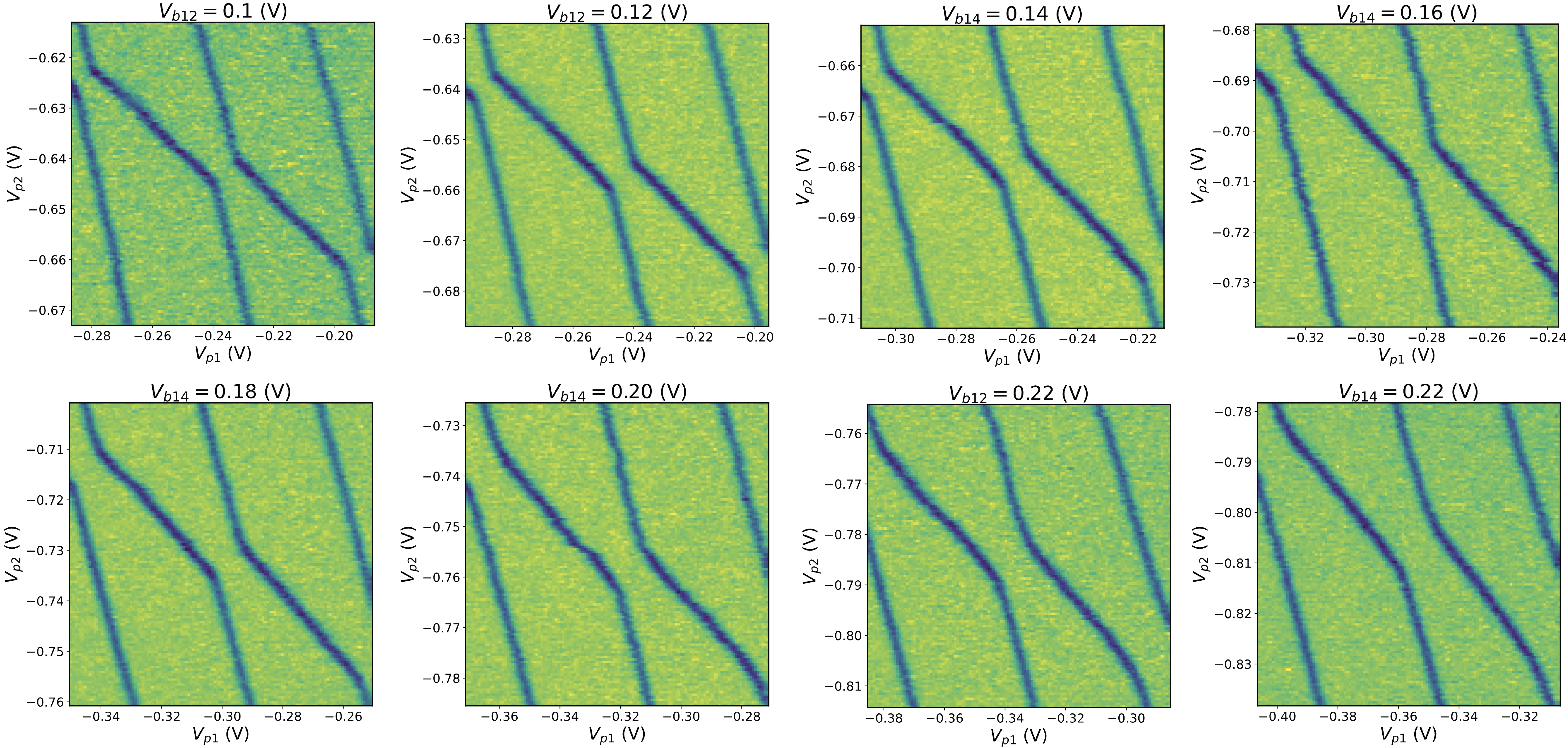}
    \caption{\label{fig2}{Anti-crossings with barrier voltage values from $0.1$ V to $0.24$ V. The window of the scan is adjusted so that every anti-crossing is roughly at the center of the scan.}}
\end{figure*}

Before the optimization, we first zoom into the target anti-crossing and perform scans with higher resolution at various barrier voltages. Specifically, we vary the barrier voltage from 0.1 V to 0.24 V. All diagrams are taken at $125\times 125$ resolution across 0.06 ($V_{p2}$) and 0.1 ($V_{p1}$) meV. These diagrams are shown in Figure \ref{fig2}. 
From these plots, we see an increase in line curvature as we increase the barrier gate voltage, suggesting an increase in tunnel coupling -- consistent to what theory predicts \cite{Wang2011}.

\begin{figure*}
    \centering
    \includegraphics[width=1.0\textwidth]{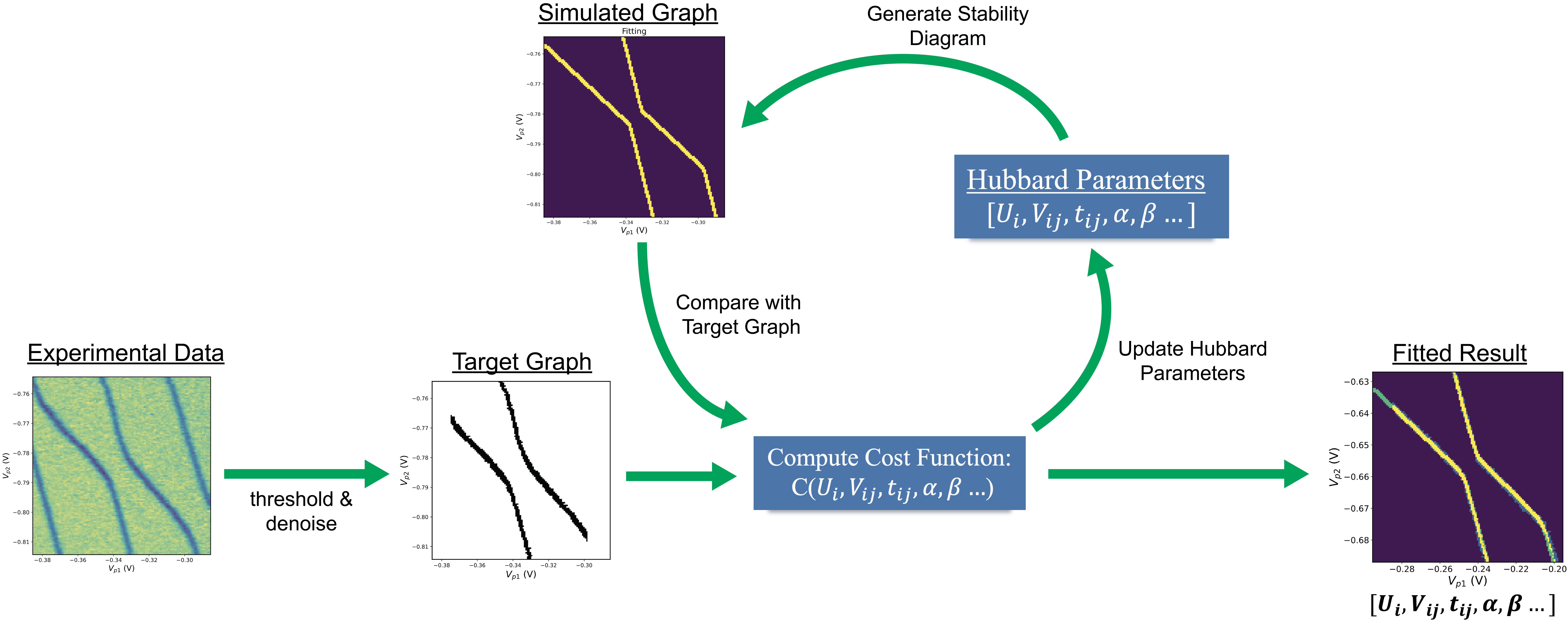}
    \caption{\label{fig3} {Optimization Demonstration. Starting from the experimental stability diagram, we first perform some thresholding and denoising to remove the unnecessary features in the stability diagram (including extra lines and corrupted regions). We also convert the z-axis values into binary values. Starting from a set of guess Hubbard parameters, we produce a simulated stability diagram in the same window and calculate a cost function that characterizes the difference between the simulated and target diagram. We then optimize the Hubbard parameters with the dual-annealing optimizer until the optimization converges. On the right, we plot the simulated diagram with the optimal parameters over the target diagram.}}
\end{figure*}

The rest of the characterization includes an optimization process that iteratively compares a simulated stability diagram and the target stability diagram. The entire optimization procedure is depicted in Figure \ref{fig3}.

We start with an experimental stability diagram. After performing a set of thresholding and denoiseing, we arrive at a binary graph without the extra features. We first generate a simulated stability diagram in the same window from a set of guess Hubbard parameters. We then compare the simulated graph with the target graph with a cost function that characterizes the difference between the two diagrams (see Supplemental Material for details). With the cost function defined, we can then perform a closed loop optimization to find the optimal value. We choose dual-annealing as the optimizer, since it has proven to be effective in escaping local minima (something we have encountered frequently). Lastly, a set of optimized Hubbard parameters is outputted after the optimization converges.

\section{Result}
In this section, we present the results of the characterization. We perform the fitting on all eight plots in the previous section with different barrier gate voltages. The fitted results are displayed in Figure 4.
\begin{figure*}
    \centering
    \includegraphics[width=1.0\textwidth]{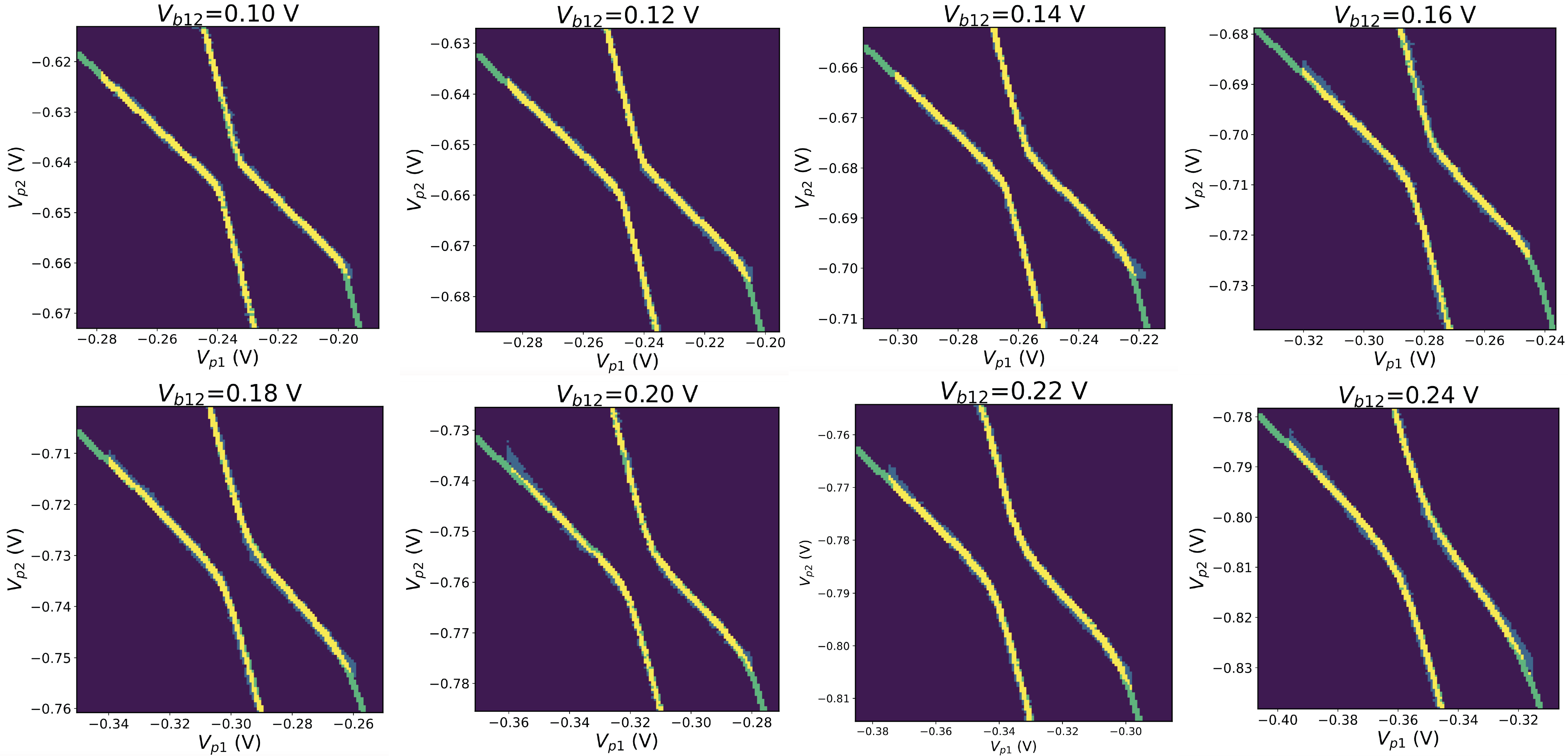}
    \caption{\label{fig4} {Fitting result of anti-crossings with various barrier gate voltages. The simulated result is plotted on top of the experimental data. Green corresponds to the simulated stability diagram; blue corresponds the experimental data; while yellow is the overlap between the two.}}
\end{figure*}

From these plots, we see good correspondence between the simulation and the experimental data. For these fitting, we assume that the cross-capacitances and the charging energies remain constant as we increase the barrier voltage, which proved to be a reasonable assumption for the regime where we performed the fitting.

\begin{figure*}
    \centering
    \includegraphics[width=1.0\textwidth]{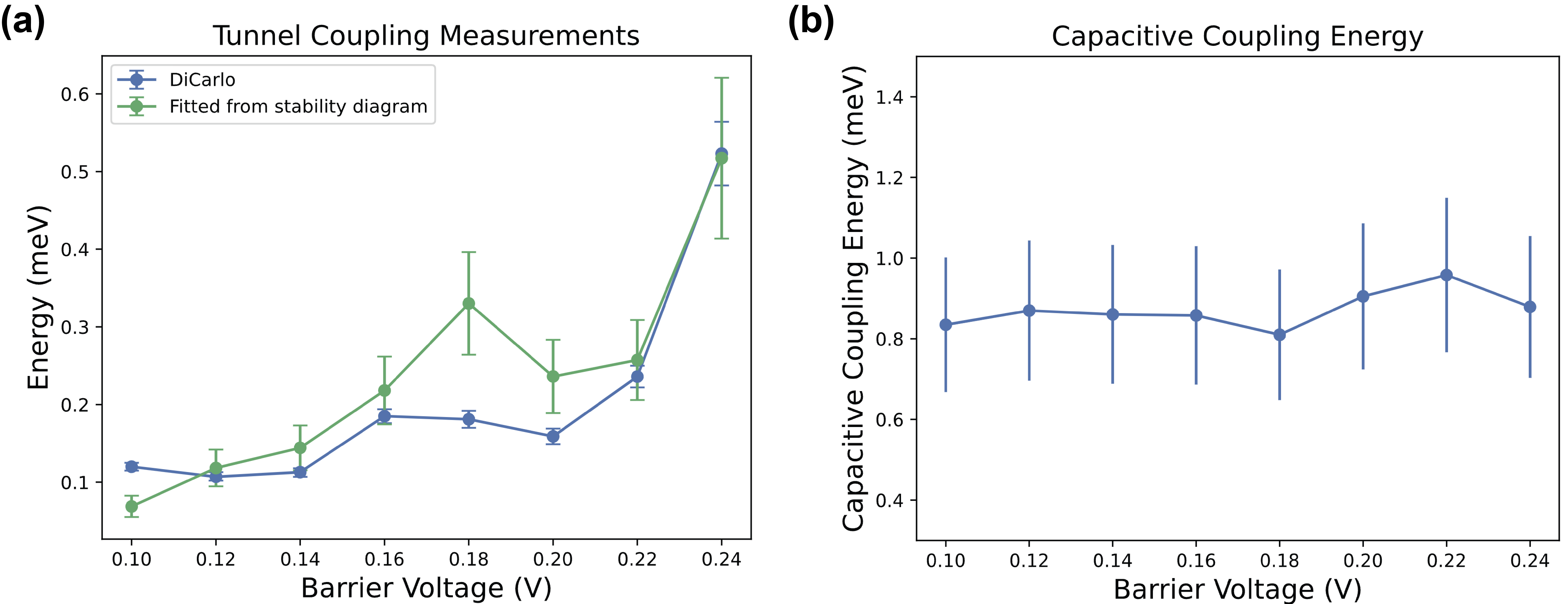}
    \caption{\label{results}{Results from the Hubbard model characterization. \textbf{(a)}: Measured tunnel couplings (green) comparison to the DiCarlo method (blue). \textbf{(b)}: Measured capacitive couplings with respect to various barrier voltages.}}
\end{figure*}

We plot the measured the capacitive couplings (or coupling energies) and the tunnel couplings in Figure \ref{results}. We observed that the coupling energies roughly remains constant in the measured region, while the tunnel coupling increases exponentially as the barrier voltage increases. In Figure \ref{results}a, we plot the measured tunnel couplings from our method to the tunnel couplings measured from the DiCarlo method (see Supplemental Material for details). We found that our results agree well with the DiCarlo method, confirming the accuracy of this method. At the time of writing this paper, since there were no known method for reliably measuring the capacitive coupling at the high tunnel coupling regime, we could not verify our electrostatic/capacitive coupling results with another known method.

We acknowledge that estimating the errors of the fitted parameters is nontrivial, since analytical solution does not exist for such a system. For the scope of this experiment, we consider the error as a result of the error in measuring lever arms. We have numerically verified that the 20\% error in lever arms results in 20\% errors in all fitted parameters. The error in the DiCarlo method is estimated from the error in FWHM from the Gaussian fitting of polarization line width.

It is important to note that there are a few limiting factors to this method. In previous experiments, we have found degeneracies while fitting many parameters, in particular the cross-capacitances and tunnel coupling. As the tunnel coupling is increased, the slopes of the transition lines between anti-crossings are corrupted by the increasing curvature; however, the cross-capacitances could ``compensate'' this effect during optimization and lead to inaccurate results. Therefore, we measured the cross-capacitances and charging energies at the low-tunnel coupling regime, and fixed their values during the optimization. As a result, this method would not be reliable if the cross-capacitance/charging energies varies much from the low-tunnel coupling regime.

Other factors include the polarization line width and stability diagram resolutions. While we found that these factors have minor effects on the results, they are not as impactful to the results as the degeneracy we have discussed previously.

\section{Conclusion}
In this work, we have developed an optimization protocol that extracts the Hubbard model parameters from the geometric features of experimental stability diagrams. Our method is able to measure large tunnel couplings, where the DiCarlo method is difficult to execute. We also demonstrated our method's ability to measure the capacitive coupling in this regime, which were only previously measured in the low tunnel coupling limit. This work allows for future studies on the relationship of the coupling between quantum dots and the barrier voltages that define their potential landscape. Furthermore, this model-based fitting approach also enables fine-tuning of the Hubbard model simulation using quantum dot arrays. We also note that this method has shown robustness to experimental noise/limitations such as missing pixels, wide polarization lines, and low pixel resolutions.

Some future work includes studying whether one could extract information regarding additional terms in the generic Hubbard model, such as the spin-exchange and pair-hopping terms. Lastly, we believe that this work can be easily generalized to a many-dot array and shows promises for high automation in the future, which is essential for the automated calibration of large quantum dot arrays.

\begin{acknowledgments}
The authors would like to thank Lisa Edge of HRL Laboratories for providing the Si/SiGe heterostructures. The work was supported by ARO through Grant. No. W911NF-23-1-0016. 
\end{acknowledgments}

\section{Appendixes}

\subsection{Stability Diagram Simulation from the Hubbard Model}
\label{SimulateSD}
It has been shown that simulated stability diagrams can be generated from the Hubbard model \cite{Yang2011}.

For a two-dot array, the Hubbard Hamiltonian with tunneling term is given:
\begin{multline}
    \cur{H} = \sum_{i=1,2}(-\mu_i\hat{n}_i+U_i\hat{n}_{i\uparrow}\hat{n}_{i\downarrow}) + U_{12}\hat{n}_1\hat{n_2} \\ - \sum_{\expt{i,j},\sigma} t(\hat{c}_{i\sigma}^\dag\hat{c}_{j\sigma} + \hat{c}_{j\sigma}^\dag\hat{c}_{i\sigma})
\end{multline}
where $\hat{n}$ is the number operator given:
\begin{align}
    \hat{n}_{i\sigma} &\equiv \hat{c}_{i\sigma}^\dag \hat{c}_{i\sigma} \\
    \hat{n}_i &\equiv \hat{n}_{i\uparrow} + \hat{n}_{i\downarrow}
\end{align}
$\mu_i$ is the chemical potential; $U_i$ and $U_{ij}$ are the on-site and off-site Coulomb interaction magnitude; and $t$ is the coupling energy between dots.

For simplicity, we assume that each site holds up to a total of two electrons - one with spin up and the other spin down. We can therefore define our basis:
\begin{equation}
    \ket{\psi} = \ket{ n_{1\uparrow},n_{2\uparrow},n_{1\downarrow},n_{2\downarrow}}
\end{equation}
Each state $\ket{n}_{i\sigma}$ represents a two level system (at site $i$, there could be $0$ or $1$ spin $\sigma$ electron), a result of the Pauli exclusion principle. Therefore, the dimension of the system is therefore $2\otimes 2 \otimes 2 \otimes 2 =16$. 

To obtain the creation and annihilation operator on site $i$, we first recall the definition of the creation and annihilation operators for a 2 level system:
\begin{equation}
    \hat{a}^\dag =
    \begin{pmatrix}
    0 & 0 \\
    1 & 0
    \end{pmatrix},
    \quad 
    \hat{a} =
    \begin{pmatrix}
    0 & 1 \\
    0 & 0
    \end{pmatrix}
\end{equation}
Now, to get the annihilation and creation operators for $\ket{n_{i\sigma}}$, we use tensor product:
\begin{align}
    \hat{c}_{i\sigma}^\dag &= \mathbb{1}_{1\uparrow}\otimes \dots \otimes \hat{a}^\dag \otimes \dots \otimes \mathbb{1}_{N\downarrow}  \\
    \hat{c}_{i\sigma} &= \mathbb{1}_{1\uparrow}\otimes \dots \otimes \hat{a} \otimes \dots \otimes \mathbb{1}_{N\downarrow}
\end{align}
where $\mathbb{1}$ is the identity matrix in $2$ dimension. To obtain the most probable configuration, we first find the eigenstates and eigenvalues of the Hamiltonian. Then, we locate the eigenstate corresponding to the lowest eigenvalue(s). That is the ground state of the Hamiltonian with a specific electron configuration. Note that the eigenstate is no longer simply a state in the standard basis for the non-zero tunneling case. After we found the eigenstate, we find the corresponding density matrix:
\begin{equation}
    \rho = \ket{\psi}\bra{\psi}
\end{equation}
Then, from the diagonal components of the matrix, we find the component of the greatest amplitude and that component will be the most probable configuration. We repeat the same procedure while varying the chemical potential on two axes. This would give us a graph a various electron configurations; to obtain a stability diagram, we take the gradient off the graph and plot the edges that separates regions with different electron configuration. The simulation result is as follows. Note that the most probable state is plotted.
\begin{figure*}
    \centering
    \includegraphics[width=0.7\textwidth]{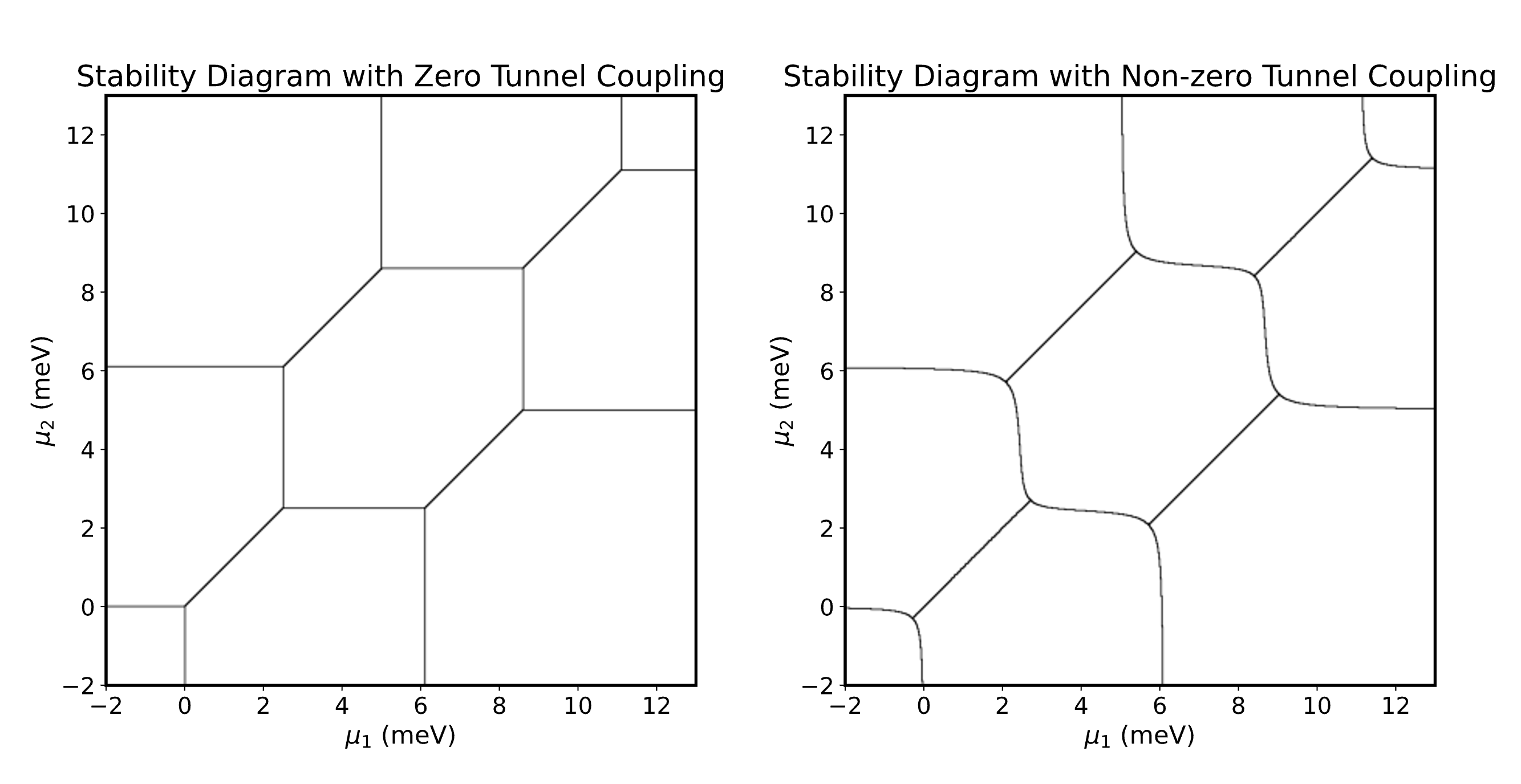}
    \caption{Simulated stability diagram from a two-level Hubbard Model. The graphs are plot in units of meV, and only the transition lines are shown in the plots. The left plot has zero tunnel coupling with parameters: $U_{1}=U_{2}=6.1 \textrm{ meV}$, $U_{12}=2.5 \textrm{ meV}$. The plot on the right has the same Hubbard parameters as the left with the exception of $t=0.3 \textrm{ meV}$. Note that the lines close to the anti-crossings are curved as a result of the tunneling effects. Further, the charging energy $U_{1(2)}$ corresponds to the horizontal/vertical spacings between the transition lines, while $U_{12}$ is correlated to the anti-crossing distance.}
\end{figure*}

The correlation between the geometry of the stability diagram and the Hubbard model parameters are well discussed in \cite{Wang2011}. The coulomb on-site energies $U_{1}$ and $U_{2}$ corresponds to the horizontal and vertical spacing between transition lines. The coulomb off-site energies energy $U_{12}$ corresponds to the diagonal distance of the anti-crossings. Lastly, the tunnel coupling $t$ curves the transition lines of further opens up the anti-crossing.

\subsection{Converting from Voltage Space to Energy Space}
\label{ConvertVE}

While the configuration of a quantum dot array can be mapped to the Hubbard model, a conversion from the voltage space to energy space is a necessity. In experiments, stability diagrams are generated by scanning through the electrical voltages of the plunger gates. In order to fit the experimental stability diagram, we need to define a relation that relates the gate voltages to the chemical potential. We use the following equations to define this relation without losing generality:
\begin{align}
\mu_{0} &= (\alpha_{1} V_{1}+\gamma_{1}) + \beta_{1}(\alpha_{2}V_{2}+\gamma_{2}) \\
\mu_{1} &=\beta_{2}(\alpha_{1} V_{1}+\gamma_{1}) + (\alpha_{2}V_{2}+\gamma_{2})
\end{align}
Note that the convention we use here is different from other references. Here, $\alpha_{i}$ is the lever arms we use to convert the voltage values to energy units, which we measure to be $0.1\textrm{eV}/\textrm{V}$ for both dots; $\gamma_{i}$ is the energy offset which depends on the specific experimental anti-crossing; finally, $\beta_{i}$ is the cross-capacitance which account for $V_{p1}$($V_{p2}$)'s effect on dot 2(1). The two equations are also easy to interpret: each chemical potential is made up by two components, one from the local dot and another from the other dot. This convention also comes in handy when we are fitting experimental diagrams, since $\beta_{i}$ corresponds to the slope (or reciprocal of the slope) of the transition lines.

\subsection{Simulation Speedup}
\label{Speedup}

While working with these simulations provides us with insight regarding the relation between the Hubbard model and stability diagrams. These simulations often take long to compute - for a single pixel on a stability diagram, we need to diagonalize a 16 by 16 matrix. If optimizations algorithms were to be used, we need to generate thousands of stability diagrams within a reasonable amount of time. Here, we propose three techniques we used to speedup the process.
\begin{enumerate}
\item Spinless Hubbard Simulation

The first method we applied is to discard the effect of spin in the Hubbard model simulation. While spin is an essential component of the Hubbard model, it simply adds degeneracies to the our simulation and modify the effective hopping term. For a two-dot system where each dot can be occupied by up to 2 electrons(either spin up or spin down), the Hilbert space is $2\otimes2\otimes2\otimes2=16$ dimensional. We can define a new basis without the consideration of spin:
\begin{equation}
\ket{\psi}=\ket{n_{1}, n_{2}}
\end{equation}
Similarly, assuming each dot holds up to $2$ electrons, the Hilbert space now has dimension $3\otimes 3=9$ instead of $16$. The Hamiltonian is now:
\begin{multline}
\cur{H} = \sum_{i=1,2}(-\mu_i\hat{n}_i+U_i\hat{n}_{i}\hat{n}_{i})  + U_{12}\hat{n}_1\hat{n_2} \\ - \sum_{\expt{i,j}} t(\hat{c}_{i}^\dag\hat{c}_{j} + \hat{c}_{j}^\dag\hat{c}_{i})
\end{multline}
where the number operators are defined: $\hat{n}_{i}=\hat{c}^{\dagger}_{i}\hat{c}_{i}$.

 However, there is a caveat to this modification. In the original Hubbard model, the hopping term implies that electron has the same spin after hopping to a different site(thus the same $\sigma$ under $\hat{c}_{i\sigma}^{\dagger}\hat{c}_{j\sigma}$). This restriction is not built into the our spinless model, and thus needs to be accounted for by other means. We notice that the effect only happens when one dot is filled with two electron of different spins and the other is filled with one. The Pauli exclusion principle only allows the electron of the opposite spin to hop to the other dot. Therefore, the hopping energy term is reduced by a factor of 2 in this particular case. In our simulation, we find the corresponding components in the matrix form of the hopping term and reduce these components by a factor of 2. The speed comparison of the two models are shown below.
 
\begin{figure}
    \centering
    \includegraphics[width=0.45\textwidth]{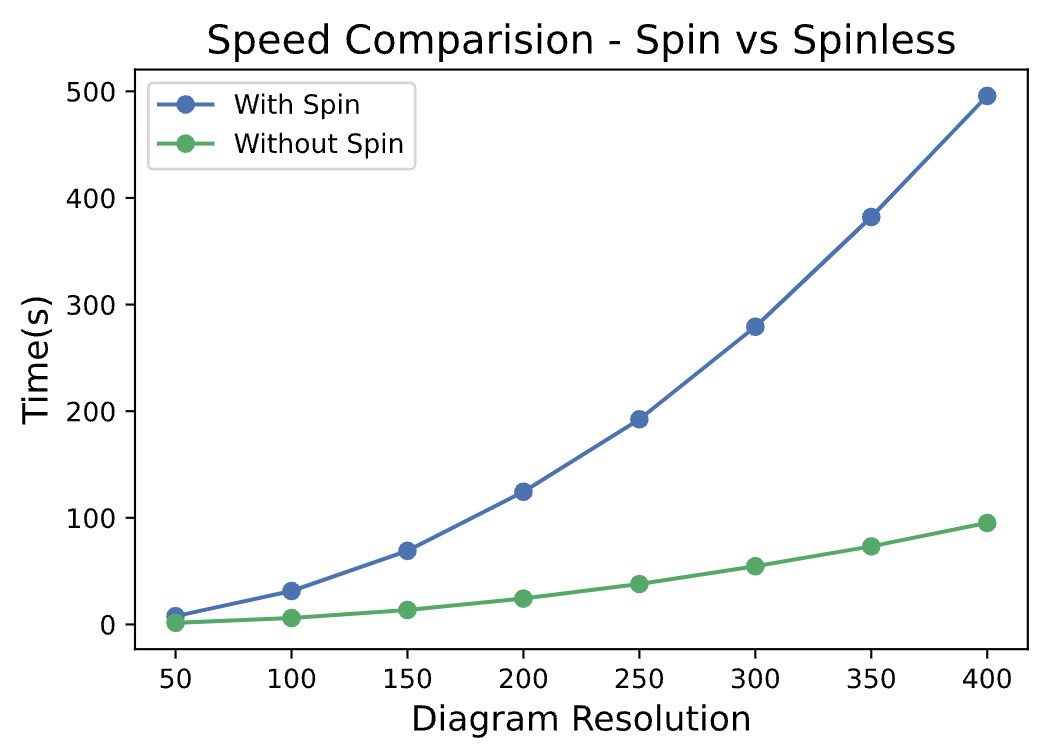}
    \caption{Speed comparison of the Hubbard model simulations with and without spin. For each model, 8 graphs with different resolutions are generated for comparison. The blue line represents the Hubbard model simulation with spin, while the green represents the spinless Hubbard model simulation.}
\end{figure}

\item Line Tracing Algorithm

To further increase the speed of the simulation, we developed a line tracing algorithm for efficiently producing stability diagrams. We first note that most of the pixels in the stability diagrams are ``blank'' and only a few pixels are ``colored'' - where transport happens. To overcome this shortcoming, we developed a line tracing algorithm that produces stability diagrams while only calculating the necessary pixels.

We notice two distinct features of any stability diagrams: 1, lines are always connected; 2, there are no isolated loops. These two features guarantee that every transition line exits the window from one of the four edges. As a result, by tracing the transition lines from the four edges, we are able to recover all transition lines in a stability diagram. The procedures of the algorithm are as follows:

\begin{enumerate}
\item Calculate the electron configurations on the four edges and identify where transition happens.
\item Store all the transition pixels' coordinates in a stack and an array.
\item For each coordinate in the stack, calculate the surrounding pixels' configurations. Add any newly discovered transition pixel's coordinate to the stack and the array. Delete the original pixel's coordinate from the stack.
\item Repeat until the stack is empty
\item Output the array consisting the transition pixels' coordinates.
\end{enumerate}
A graphical demonstration is shown in Figure \ref{lineTrace}:
\begin{figure*}
    \centering
    \includegraphics[width=0.6\textwidth]{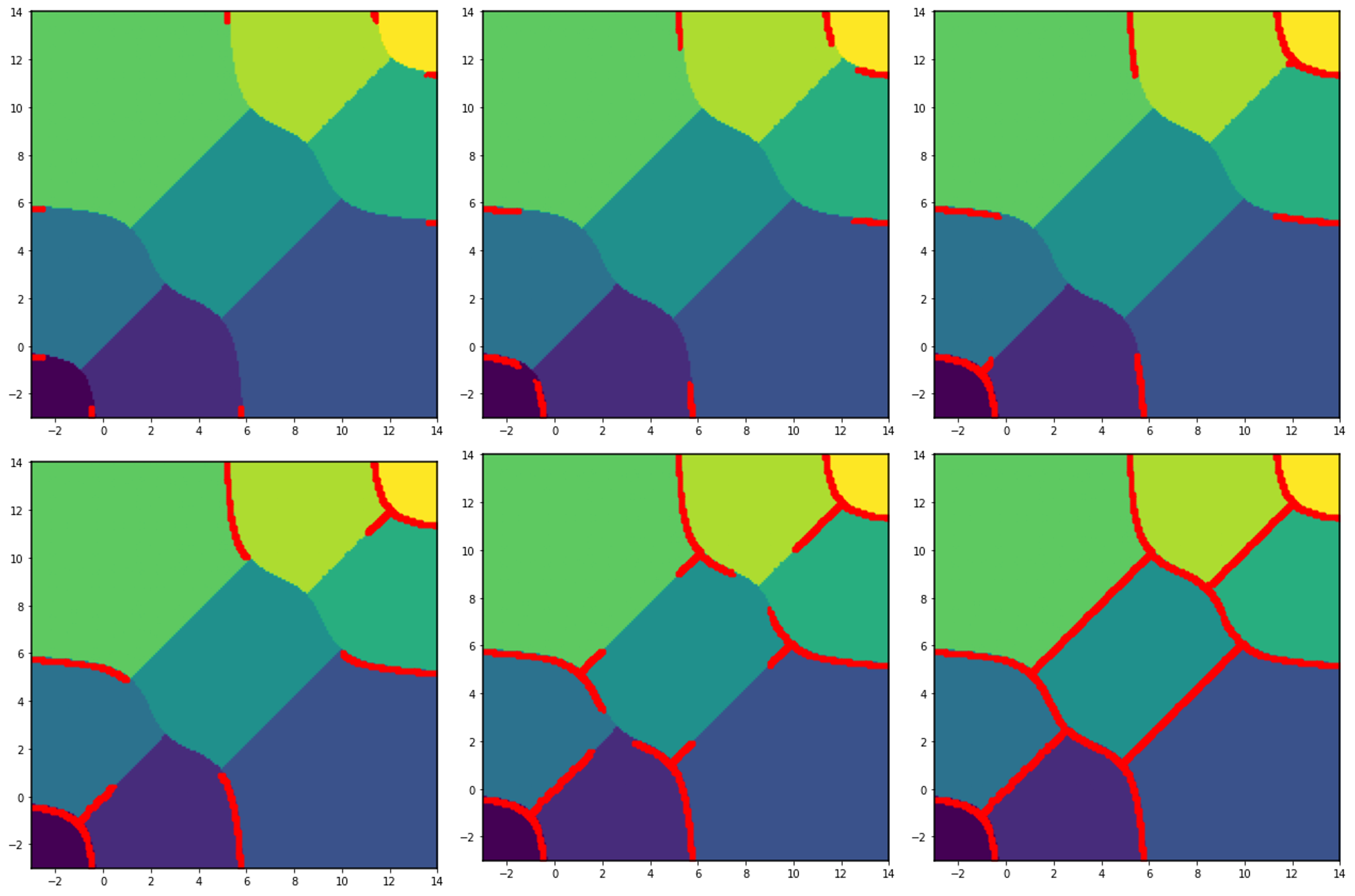}
    \caption{These six graphs provide a visualization of the line-tracing algorithm. The red pixels demonstrate the pixels that were added to the stack. The algorithm starts from the edges(first figure) and looks for transition pixels from neighboring pixels. We see that all the transitions are identified in the last graph.}
\label{lineTrace}
\end{figure*}
A speed comparison between the line tracing and without the line tracing algorithm is shown in Figure \ref{SpeedComparision2}:

\end{enumerate}
\begin{figure}
    \centering
    \includegraphics[width=0.45\textwidth]{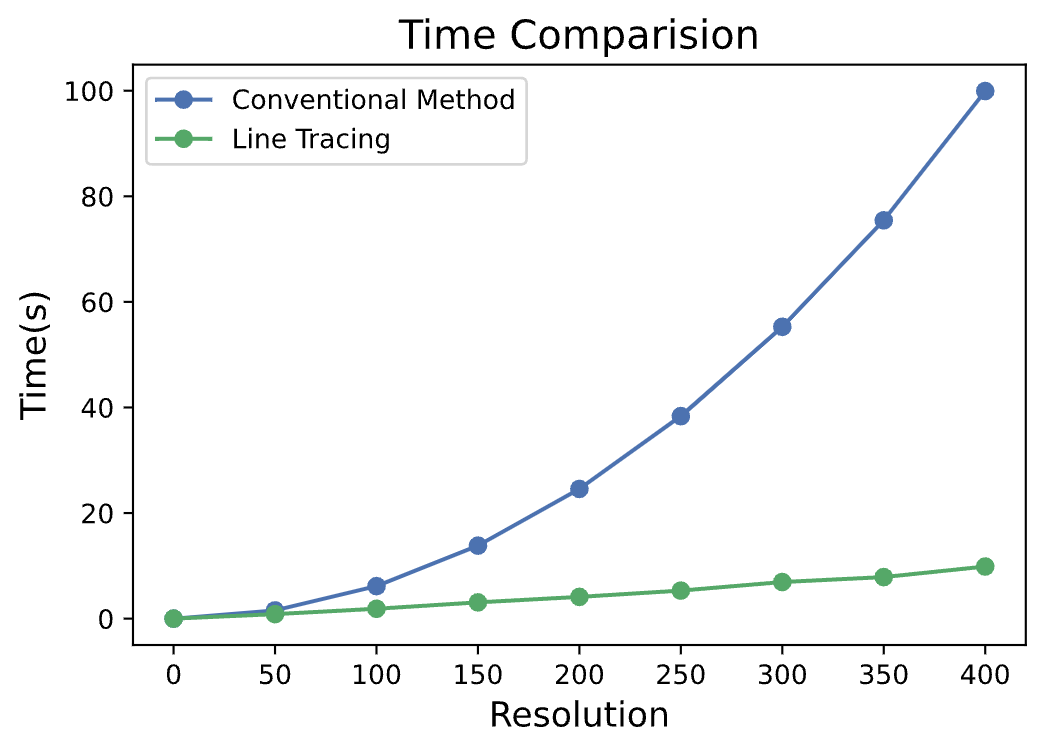}
    \caption{A time comparison between the Hubbard model simulation with and without the line tracing algorithm. Far less pixels are calculated for the line tracing algorithm, and we observe an exponential speedup as the diagram resolution increases.}
\label{SpeedComparision2}
\end{figure}

\subsection{Cost Function Definition}

In this section, we will discuss how we defined the cost function when comparing the target and simulated stability diagrams. We have previously tried standard methods in computer vision for comparing images, such as the mean squared error and the structural similarity measure. However, we have found them to be ineffective due to the binary nature of the stability diagrams, and the single pixel sensitivity in our fittings. Therefore, we defined the cost function using the binary nature of the stability diagrams, while maintaining sensitivity to error in single pixels.

Suppose we have a target (experimental) and simulated stability diagrams, and they are stored in matrices where 0 depicts background and 1 depicts transitions. The cost function is defined as follows:
\begin{enumerate}
\item Find all ``1'' pixels in the simulated stability diagram, and iterate through the list.
\item For each ``1'' pixel, we record its coordinates and look at the pixel of the same coordinates in the target stability diagram.
\item At this pixel, find the distance to the nearest ``1'' pixel in the target stability diagram. If no ``1'' pixel was found, use the dimension/2 of the stability diagram as distance.
\item Record this distance, and repeat this process for all ``1'' pixels in the simulated diagram.
\item Output the sum of the distances.
\end{enumerate}
Therefore, the distance sum would characterize this difference between the two diagrams. It is obvious that if the two diagrams are the same, this total distance would be 0.

% The \nocite command causes all entries in a bibliography to be printed out
% whether or not they are actually referenced in the text. This is appropriate
% for the sample file to show the different styles of references, but authors
% most likely will not want to use it.
\nocite{*}

\bibliography{apssamp}% Produces the bibliography via BibTeX.

%apsrev4-2.bst 2019-01-14 (MD) hand-edited version of apsrev4-1.bst
%Control: key (0)
%Control: author (8) initials jnrlst
%Control: editor formatted (1) identically to author
%Control: production of article title (0) allowed
%Control: page (0) single
%Control: year (1) truncated
%Control: production of eprint (0) enabled
\begin{thebibliography}{16}%
\makeatletter
\providecommand \@ifxundefined [1]{%
 \@ifx{#1\undefined}
}%
\providecommand \@ifnum [1]{%
 \ifnum #1\expandafter \@firstoftwo
 \else \expandafter \@secondoftwo
 \fi
}%
\providecommand \@ifx [1]{%
 \ifx #1\expandafter \@firstoftwo
 \else \expandafter \@secondoftwo
 \fi
}%
\providecommand \natexlab [1]{#1}%
\providecommand \enquote  [1]{``#1''}%
\providecommand \bibnamefont  [1]{#1}%
\providecommand \bibfnamefont [1]{#1}%
\providecommand \citenamefont [1]{#1}%
\providecommand \href@noop [0]{\@secondoftwo}%
\providecommand \href [0]{\begingroup \@sanitize@url \@href}%
\providecommand \@href[1]{\@@startlink{#1}\@@href}%
\providecommand \@@href[1]{\endgroup#1\@@endlink}%
\providecommand \@sanitize@url [0]{\catcode `\\12\catcode `\$12\catcode
  `\&12\catcode `\#12\catcode `\^12\catcode `\_12\catcode `\%12\relax}%
\providecommand \@@startlink[1]{}%
\providecommand \@@endlink[0]{}%
\providecommand \url  [0]{\begingroup\@sanitize@url \@url }%
\providecommand \@url [1]{\endgroup\@href {#1}{\urlprefix }}%
\providecommand \urlprefix  [0]{URL }%
\providecommand \Eprint [0]{\href }%
\providecommand \doibase [0]{https://doi.org/}%
\providecommand \selectlanguage [0]{\@gobble}%
\providecommand \bibinfo  [0]{\@secondoftwo}%
\providecommand \bibfield  [0]{\@secondoftwo}%
\providecommand \translation [1]{[#1]}%
\providecommand \BibitemOpen [0]{}%
\providecommand \bibitemStop [0]{}%
\providecommand \bibitemNoStop [0]{.\EOS\space}%
\providecommand \EOS [0]{\spacefactor3000\relax}%
\providecommand \BibitemShut  [1]{\csname bibitem#1\endcsname}%
\let\auto@bib@innerbib\@empty
%</preamble>
\bibitem [{\citenamefont {Sarma}\ \emph {et~al.}(2011)\citenamefont {Sarma},
  \citenamefont {Wang},\ and\ \citenamefont {Yang}}]{Sarma2011}%
  \BibitemOpen
  \bibfield  {author} {\bibinfo {author} {\bibfnamefont {S.~D.}\ \bibnamefont
  {Sarma}}, \bibinfo {author} {\bibfnamefont {X.}~\bibnamefont {Wang}},\ and\
  \bibinfo {author} {\bibfnamefont {S.}~\bibnamefont {Yang}},\ }\bibfield
  {title} {\bibinfo {title} {Hubbard model description of silicon spin qubits:
  Charge stability diagram and tunnel coupling in si double quantum dots},\
  }\bibfield  {journal} {\bibinfo  {journal} {Physical Review B - Condensed
  Matter and Materials Physics}\ }\textbf {\bibinfo {volume} {83}},\ \href
  {https://doi.org/10.1103/PhysRevB.83.235314} {10.1103/PhysRevB.83.235314}
  (\bibinfo {year} {2011})\BibitemShut {NoStop}%
\bibitem [{\citenamefont {Wang}\ \emph {et~al.}(2011)\citenamefont {Wang},
  \citenamefont {Yang},\ and\ \citenamefont {Sarma}}]{Wang2011}%
  \BibitemOpen
  \bibfield  {author} {\bibinfo {author} {\bibfnamefont {X.}~\bibnamefont
  {Wang}}, \bibinfo {author} {\bibfnamefont {S.}~\bibnamefont {Yang}},\ and\
  \bibinfo {author} {\bibfnamefont {S.~D.}\ \bibnamefont {Sarma}},\ }\bibfield
  {title} {\bibinfo {title} {Quantum theory of the charge-stability diagram of
  semiconductor double-quantum-dot systems},\ }\bibfield  {journal} {\bibinfo
  {journal} {Physical Review B - Condensed Matter and Materials Physics}\
  }\textbf {\bibinfo {volume} {84}},\ \href
  {https://doi.org/10.1103/PhysRevB.84.115301} {10.1103/PhysRevB.84.115301}
  (\bibinfo {year} {2011})\BibitemShut {NoStop}%
\bibitem [{\citenamefont {Yang}\ \emph {et~al.}(2011)\citenamefont {Yang},
  \citenamefont {Wang},\ and\ \citenamefont {Sarma}}]{Yang2011}%
  \BibitemOpen
  \bibfield  {author} {\bibinfo {author} {\bibfnamefont {S.}~\bibnamefont
  {Yang}}, \bibinfo {author} {\bibfnamefont {X.}~\bibnamefont {Wang}},\ and\
  \bibinfo {author} {\bibfnamefont {S.~D.}\ \bibnamefont {Sarma}},\ }\bibfield
  {title} {\bibinfo {title} {Generic hubbard model description of semiconductor
  quantum-dot spin qubits},\ }\bibfield  {journal} {\bibinfo  {journal}
  {Physical Review B - Condensed Matter and Materials Physics}\ }\textbf
  {\bibinfo {volume} {83}},\ \href {https://doi.org/10.1103/PhysRevB.83.161301}
  {10.1103/PhysRevB.83.161301} (\bibinfo {year} {2011})\BibitemShut {NoStop}%
\bibitem [{\citenamefont {Wei}\ \emph {et~al.}(2013)\citenamefont {Wei},
  \citenamefont {Li}, \citenamefont {Cao}, \citenamefont {Luo}, \citenamefont
  {Zheng}, \citenamefont {Tu}, \citenamefont {Xiao}, \citenamefont {Guo},
  \citenamefont {Jiang},\ and\ \citenamefont {Guo}}]{Wei2013}%
  \BibitemOpen
  \bibfield  {author} {\bibinfo {author} {\bibfnamefont {D.}~\bibnamefont
  {Wei}}, \bibinfo {author} {\bibfnamefont {H.~O.}\ \bibnamefont {Li}},
  \bibinfo {author} {\bibfnamefont {G.}~\bibnamefont {Cao}}, \bibinfo {author}
  {\bibfnamefont {G.}~\bibnamefont {Luo}}, \bibinfo {author} {\bibfnamefont
  {Z.~X.}\ \bibnamefont {Zheng}}, \bibinfo {author} {\bibfnamefont
  {T.}~\bibnamefont {Tu}}, \bibinfo {author} {\bibfnamefont {M.}~\bibnamefont
  {Xiao}}, \bibinfo {author} {\bibfnamefont {G.~C.}\ \bibnamefont {Guo}},
  \bibinfo {author} {\bibfnamefont {H.~W.}\ \bibnamefont {Jiang}},\ and\
  \bibinfo {author} {\bibfnamefont {G.~P.}\ \bibnamefont {Guo}},\ }\bibfield
  {title} {\bibinfo {title} {Tuning inter-dot tunnel coupling of an etched
  graphene double quantum dot by adjacent metal gates},\ }\bibfield  {journal}
  {\bibinfo  {journal} {Scientific Reports}\ }\textbf {\bibinfo {volume} {3}},\
  \href {https://doi.org/10.1038/srep03175} {10.1038/srep03175} (\bibinfo
  {year} {2013})\BibitemShut {NoStop}%
\bibitem [{\citenamefont {Mills}\ \emph {et~al.}(2022)\citenamefont {Mills},
  \citenamefont {Guinn}, \citenamefont {Gullans}, \citenamefont {Sigillito},
  \citenamefont {Feldman}, \citenamefont {Nielsen},\ and\ \citenamefont
  {Petta}}]{Mills2022}%
  \BibitemOpen
  \bibfield  {author} {\bibinfo {author} {\bibfnamefont {A.~R.}\ \bibnamefont
  {Mills}}, \bibinfo {author} {\bibfnamefont {C.~R.}\ \bibnamefont {Guinn}},
  \bibinfo {author} {\bibfnamefont {M.~J.}\ \bibnamefont {Gullans}}, \bibinfo
  {author} {\bibfnamefont {A.~J.}\ \bibnamefont {Sigillito}}, \bibinfo {author}
  {\bibfnamefont {M.~M.}\ \bibnamefont {Feldman}}, \bibinfo {author}
  {\bibfnamefont {E.}~\bibnamefont {Nielsen}},\ and\ \bibinfo {author}
  {\bibfnamefont {J.~R.}\ \bibnamefont {Petta}},\ }\bibfield  {title} {\bibinfo
  {title} {Two-qubit silicon quantum processor with operation fidelity
  exceeding 99\%},\ }\href {https://www.science.org} {\bibfield  {journal}
  {\bibinfo  {journal} {Sci. Adv}\ }\textbf {\bibinfo {volume} {8}},\ \bibinfo
  {pages} {5130} (\bibinfo {year} {2022})}\BibitemShut {NoStop}%
\bibitem [{\citenamefont {Yang}\ \emph {et~al.}(2017)\citenamefont {Yang},
  \citenamefont {Coppersmith},\ and\ \citenamefont {Friesen}}]{Yang2017}%
  \BibitemOpen
  \bibfield  {author} {\bibinfo {author} {\bibfnamefont {Y.~C.}\ \bibnamefont
  {Yang}}, \bibinfo {author} {\bibfnamefont {S.~N.}\ \bibnamefont
  {Coppersmith}},\ and\ \bibinfo {author} {\bibfnamefont {M.}~\bibnamefont
  {Friesen}},\ }\bibfield  {title} {\bibinfo {title} {Achieving high-fidelity
  single-qubit gates in a strongly driven silicon-quantum-dot hybrid qubit},\
  }\bibfield  {journal} {\bibinfo  {journal} {Physical Review A}\ }\textbf
  {\bibinfo {volume} {95}},\ \href {https://doi.org/10.1103/PhysRevA.95.062321}
  {10.1103/PhysRevA.95.062321} (\bibinfo {year} {2017})\BibitemShut {NoStop}%
\bibitem [{\citenamefont {Yoneda}\ \emph {et~al.}(2018)\citenamefont {Yoneda},
  \citenamefont {Takeda}, \citenamefont {Otsuka}, \citenamefont {Nakajima},
  \citenamefont {Delbecq}, \citenamefont {Allison}, \citenamefont {Honda},
  \citenamefont {Kodera}, \citenamefont {Oda}, \citenamefont {Hoshi},
  \citenamefont {Usami}, \citenamefont {Itoh},\ and\ \citenamefont
  {Tarucha}}]{Yoneda2018}%
  \BibitemOpen
  \bibfield  {author} {\bibinfo {author} {\bibfnamefont {J.}~\bibnamefont
  {Yoneda}}, \bibinfo {author} {\bibfnamefont {K.}~\bibnamefont {Takeda}},
  \bibinfo {author} {\bibfnamefont {T.}~\bibnamefont {Otsuka}}, \bibinfo
  {author} {\bibfnamefont {T.}~\bibnamefont {Nakajima}}, \bibinfo {author}
  {\bibfnamefont {M.~R.}\ \bibnamefont {Delbecq}}, \bibinfo {author}
  {\bibfnamefont {G.}~\bibnamefont {Allison}}, \bibinfo {author} {\bibfnamefont
  {T.}~\bibnamefont {Honda}}, \bibinfo {author} {\bibfnamefont
  {T.}~\bibnamefont {Kodera}}, \bibinfo {author} {\bibfnamefont
  {S.}~\bibnamefont {Oda}}, \bibinfo {author} {\bibfnamefont {Y.}~\bibnamefont
  {Hoshi}}, \bibinfo {author} {\bibfnamefont {N.}~\bibnamefont {Usami}},
  \bibinfo {author} {\bibfnamefont {K.~M.}\ \bibnamefont {Itoh}},\ and\
  \bibinfo {author} {\bibfnamefont {S.}~\bibnamefont {Tarucha}},\ }\bibfield
  {title} {\bibinfo {title} {A quantum-dot spin qubit with coherence limited by
  charge noise and fidelity higher than 99.9\%},\ }\href
  {https://doi.org/10.1038/s41565-017-0014-x} {\bibfield  {journal} {\bibinfo
  {journal} {Nature Nanotechnology}\ }\textbf {\bibinfo {volume} {13}},\
  \bibinfo {pages} {102} (\bibinfo {year} {2018})}\BibitemShut {NoStop}%
\bibitem [{\citenamefont {Hensgens}\ \emph {et~al.}(2017)\citenamefont
  {Hensgens}, \citenamefont {Fujita}, \citenamefont {Janssen}, \citenamefont
  {Li}, \citenamefont {Diepen}, \citenamefont {Reichl}, \citenamefont
  {Wegscheider}, \citenamefont {Sarma},\ and\ \citenamefont
  {Vandersypen}}]{Hensgens2017}%
  \BibitemOpen
  \bibfield  {author} {\bibinfo {author} {\bibfnamefont {T.}~\bibnamefont
  {Hensgens}}, \bibinfo {author} {\bibfnamefont {T.}~\bibnamefont {Fujita}},
  \bibinfo {author} {\bibfnamefont {L.}~\bibnamefont {Janssen}}, \bibinfo
  {author} {\bibfnamefont {X.}~\bibnamefont {Li}}, \bibinfo {author}
  {\bibfnamefont {C.~J.~V.}\ \bibnamefont {Diepen}}, \bibinfo {author}
  {\bibfnamefont {C.}~\bibnamefont {Reichl}}, \bibinfo {author} {\bibfnamefont
  {W.}~\bibnamefont {Wegscheider}}, \bibinfo {author} {\bibfnamefont {S.~D.}\
  \bibnamefont {Sarma}},\ and\ \bibinfo {author} {\bibfnamefont {L.~M.}\
  \bibnamefont {Vandersypen}},\ }\bibfield  {title} {\bibinfo {title} {Quantum
  simulation of a fermi-hubbard model using a semiconductor quantum dot
  array},\ }\href {https://doi.org/10.1038/nature23022} {\bibfield  {journal}
  {\bibinfo  {journal} {Nature}\ }\textbf {\bibinfo {volume} {548}},\ \bibinfo
  {pages} {70} (\bibinfo {year} {2017})}\BibitemShut {NoStop}%
\bibitem [{\citenamefont {DiCarlo}\ \emph {et~al.}(2004)\citenamefont
  {DiCarlo}, \citenamefont {Lynch}, \citenamefont {Johnson}, \citenamefont
  {Childress}, \citenamefont {Crockett}, \citenamefont {Marcus}, \citenamefont
  {Hanson},\ and\ \citenamefont {Gossard}}]{DiCarlo2004}%
  \BibitemOpen
  \bibfield  {author} {\bibinfo {author} {\bibfnamefont {L.}~\bibnamefont
  {DiCarlo}}, \bibinfo {author} {\bibfnamefont {H.~J.}\ \bibnamefont {Lynch}},
  \bibinfo {author} {\bibfnamefont {A.~C.}\ \bibnamefont {Johnson}}, \bibinfo
  {author} {\bibfnamefont {L.~I.}\ \bibnamefont {Childress}}, \bibinfo {author}
  {\bibfnamefont {K.}~\bibnamefont {Crockett}}, \bibinfo {author}
  {\bibfnamefont {C.~M.}\ \bibnamefont {Marcus}}, \bibinfo {author}
  {\bibfnamefont {M.~P.}\ \bibnamefont {Hanson}},\ and\ \bibinfo {author}
  {\bibfnamefont {A.~C.}\ \bibnamefont {Gossard}},\ }\bibfield  {title}
  {\bibinfo {title} {Differential charge sensing and charge delocalization in a
  tunable double quantum dot},\ }\href
  {https://doi.org/10.1103/PhysRevLett.92.226801} {\bibfield  {journal}
  {\bibinfo  {journal} {Phys. Rev. Lett.}\ }\textbf {\bibinfo {volume} {92}},\
  \bibinfo {pages} {226801} (\bibinfo {year} {2004})}\BibitemShut {NoStop}%
\bibitem [{\citenamefont {Neyens}\ \emph {et~al.}(2019)\citenamefont {Neyens},
  \citenamefont {MacQuarrie}, \citenamefont {Dodson}, \citenamefont {Corrigan},
  \citenamefont {Holman}, \citenamefont {Thorgrimsson}, \citenamefont {Palma},
  \citenamefont {McJunkin}, \citenamefont {Edge}, \citenamefont {Friesen},
  \citenamefont {Coppersmith},\ and\ \citenamefont {Eriksson}}]{Neyens2019}%
  \BibitemOpen
  \bibfield  {author} {\bibinfo {author} {\bibfnamefont {S.~F.}\ \bibnamefont
  {Neyens}}, \bibinfo {author} {\bibfnamefont {E.~R.}\ \bibnamefont
  {MacQuarrie}}, \bibinfo {author} {\bibfnamefont {J.~P.}\ \bibnamefont
  {Dodson}}, \bibinfo {author} {\bibfnamefont {J.}~\bibnamefont {Corrigan}},
  \bibinfo {author} {\bibfnamefont {N.}~\bibnamefont {Holman}}, \bibinfo
  {author} {\bibfnamefont {B.}~\bibnamefont {Thorgrimsson}}, \bibinfo {author}
  {\bibfnamefont {M.}~\bibnamefont {Palma}}, \bibinfo {author} {\bibfnamefont
  {T.}~\bibnamefont {McJunkin}}, \bibinfo {author} {\bibfnamefont {L.~F.}\
  \bibnamefont {Edge}}, \bibinfo {author} {\bibfnamefont {M.}~\bibnamefont
  {Friesen}}, \bibinfo {author} {\bibfnamefont {S.~N.}\ \bibnamefont
  {Coppersmith}},\ and\ \bibinfo {author} {\bibfnamefont {M.~A.}\ \bibnamefont
  {Eriksson}},\ }\bibfield  {title} {\bibinfo {title} {Measurements of
  capacitive coupling within a quadruple-quantum-dot array},\ }\bibfield
  {journal} {\bibinfo  {journal} {Physical Review Applied}\ }\textbf {\bibinfo
  {volume} {12}},\ \href {https://doi.org/10.1103/PhysRevApplied.12.064049}
  {10.1103/PhysRevApplied.12.064049} (\bibinfo {year} {2019})\BibitemShut
  {NoStop}%
\bibitem [{\citenamefont {Diepen}\ \emph {et~al.}(2018)\citenamefont {Diepen},
  \citenamefont {Eendebak}, \citenamefont {Buijtendorp}, \citenamefont
  {Mukhopadhyay}, \citenamefont {Fujita}, \citenamefont {Reichl}, \citenamefont
  {Wegscheider},\ and\ \citenamefont {Vandersypen}}]{VanDiepen}%
  \BibitemOpen
  \bibfield  {author} {\bibinfo {author} {\bibfnamefont {C.~J.~V.}\
  \bibnamefont {Diepen}}, \bibinfo {author} {\bibfnamefont {P.~T.}\
  \bibnamefont {Eendebak}}, \bibinfo {author} {\bibfnamefont {B.~T.}\
  \bibnamefont {Buijtendorp}}, \bibinfo {author} {\bibfnamefont
  {U.}~\bibnamefont {Mukhopadhyay}}, \bibinfo {author} {\bibfnamefont
  {T.}~\bibnamefont {Fujita}}, \bibinfo {author} {\bibfnamefont
  {C.}~\bibnamefont {Reichl}}, \bibinfo {author} {\bibfnamefont
  {W.}~\bibnamefont {Wegscheider}},\ and\ \bibinfo {author} {\bibfnamefont
  {L.~M.}\ \bibnamefont {Vandersypen}},\ }\bibfield  {title} {\bibinfo {title}
  {Automated tuning of inter-dot tunnel coupling in double quantum dots},\
  }\bibfield  {journal} {\bibinfo  {journal} {Applied Physics Letters}\
  }\textbf {\bibinfo {volume} {113}},\ \href
  {https://doi.org/10.1063/1.5031034} {10.1063/1.5031034} (\bibinfo {year}
  {2018})\BibitemShut {NoStop}%
\bibitem [{\citenamefont {Nurizzo}\ \emph {et~al.}(2022)\citenamefont
  {Nurizzo}, \citenamefont {Jadot}, \citenamefont {Mortemousque}, \citenamefont
  {Thiney}, \citenamefont {Chanrion}, \citenamefont {Dartiailh}, \citenamefont
  {Ludwig}, \citenamefont {Wieck}, \citenamefont {Bäuerle}, \citenamefont
  {Urdampilleta},\ and\ \citenamefont {Meunier}}]{Nurizzo2022}%
  \BibitemOpen
  \bibfield  {author} {\bibinfo {author} {\bibfnamefont {M.}~\bibnamefont
  {Nurizzo}}, \bibinfo {author} {\bibfnamefont {B.}~\bibnamefont {Jadot}},
  \bibinfo {author} {\bibfnamefont {P.~A.}\ \bibnamefont {Mortemousque}},
  \bibinfo {author} {\bibfnamefont {V.}~\bibnamefont {Thiney}}, \bibinfo
  {author} {\bibfnamefont {E.}~\bibnamefont {Chanrion}}, \bibinfo {author}
  {\bibfnamefont {M.}~\bibnamefont {Dartiailh}}, \bibinfo {author}
  {\bibfnamefont {A.}~\bibnamefont {Ludwig}}, \bibinfo {author} {\bibfnamefont
  {A.~D.}\ \bibnamefont {Wieck}}, \bibinfo {author} {\bibfnamefont
  {C.}~\bibnamefont {Bäuerle}}, \bibinfo {author} {\bibfnamefont
  {M.}~\bibnamefont {Urdampilleta}},\ and\ \bibinfo {author} {\bibfnamefont
  {T.}~\bibnamefont {Meunier}},\ }\bibfield  {title} {\bibinfo {title}
  {Controlled quantum dot array segmentation via highly tunable interdot tunnel
  coupling},\ }\bibfield  {journal} {\bibinfo  {journal} {Applied Physics
  Letters}\ }\textbf {\bibinfo {volume} {121}},\ \href
  {https://doi.org/10.1063/5.0105635} {10.1063/5.0105635} (\bibinfo {year}
  {2022})\BibitemShut {NoStop}%
\bibitem [{\citenamefont {Liu}\ \emph {et~al.}(2022)\citenamefont {Liu},
  \citenamefont {Wang}, \citenamefont {Wang}, \citenamefont {Sun},
  \citenamefont {Yin}, \citenamefont {Li}, \citenamefont {Cao},\ and\
  \citenamefont {Guo}}]{Liu2022}%
  \BibitemOpen
  \bibfield  {author} {\bibinfo {author} {\bibfnamefont {H.}~\bibnamefont
  {Liu}}, \bibinfo {author} {\bibfnamefont {B.}~\bibnamefont {Wang}}, \bibinfo
  {author} {\bibfnamefont {N.}~\bibnamefont {Wang}}, \bibinfo {author}
  {\bibfnamefont {Z.}~\bibnamefont {Sun}}, \bibinfo {author} {\bibfnamefont
  {H.}~\bibnamefont {Yin}}, \bibinfo {author} {\bibfnamefont {H.}~\bibnamefont
  {Li}}, \bibinfo {author} {\bibfnamefont {G.}~\bibnamefont {Cao}},\ and\
  \bibinfo {author} {\bibfnamefont {G.}~\bibnamefont {Guo}},\ }\bibfield
  {title} {\bibinfo {title} {An automated approach for consecutive tuning of
  quantum dot arrays},\ }\bibfield  {journal} {\bibinfo  {journal} {Applied
  Physics Letters}\ }\textbf {\bibinfo {volume} {121}},\ \href
  {https://doi.org/10.1063/5.0111128} {10.1063/5.0111128} (\bibinfo {year}
  {2022})\BibitemShut {NoStop}%
\bibitem [{\citenamefont {Van~der Wiel}\ \emph {et~al.}(2002)\citenamefont
  {Van~der Wiel}, \citenamefont {De~Franceschi}, \citenamefont {Elzerman},
  \citenamefont {Fujisawa}, \citenamefont {Tarucha},\ and\ \citenamefont
  {Kouwenhoven}}]{VanDerWiel}%
  \BibitemOpen
  \bibfield  {author} {\bibinfo {author} {\bibfnamefont {W.~G.}\ \bibnamefont
  {Van~der Wiel}}, \bibinfo {author} {\bibfnamefont {S.}~\bibnamefont
  {De~Franceschi}}, \bibinfo {author} {\bibfnamefont {J.~M.}\ \bibnamefont
  {Elzerman}}, \bibinfo {author} {\bibfnamefont {T.}~\bibnamefont {Fujisawa}},
  \bibinfo {author} {\bibfnamefont {S.}~\bibnamefont {Tarucha}},\ and\ \bibinfo
  {author} {\bibfnamefont {L.~P.}\ \bibnamefont {Kouwenhoven}},\ }\bibfield
  {title} {\bibinfo {title} {Electron transport through double quantum dots},\
  }\href@noop {} {\bibfield  {journal} {\bibinfo  {journal} {Reviews of modern
  physics}\ }\textbf {\bibinfo {volume} {75}},\ \bibinfo {pages} {1} (\bibinfo
  {year} {2002})}\BibitemShut {NoStop}%
\bibitem [{\citenamefont {Volk}\ \emph {et~al.}(2019)\citenamefont {Volk},
  \citenamefont {Zwerver}, \citenamefont {Mukhopadhyay}, \citenamefont
  {Eendebak}, \citenamefont {van Diepen}, \citenamefont {Dehollain},
  \citenamefont {Hensgens}, \citenamefont {Fujita}, \citenamefont {Reichl},
  \citenamefont {Wegscheider},\ and\ \citenamefont {Vandersypen}}]{Volk2019}%
  \BibitemOpen
  \bibfield  {author} {\bibinfo {author} {\bibfnamefont {C.}~\bibnamefont
  {Volk}}, \bibinfo {author} {\bibfnamefont {A.~M.}\ \bibnamefont {Zwerver}},
  \bibinfo {author} {\bibfnamefont {U.}~\bibnamefont {Mukhopadhyay}}, \bibinfo
  {author} {\bibfnamefont {P.~T.}\ \bibnamefont {Eendebak}}, \bibinfo {author}
  {\bibfnamefont {C.~J.}\ \bibnamefont {van Diepen}}, \bibinfo {author}
  {\bibfnamefont {J.~P.}\ \bibnamefont {Dehollain}}, \bibinfo {author}
  {\bibfnamefont {T.}~\bibnamefont {Hensgens}}, \bibinfo {author}
  {\bibfnamefont {T.}~\bibnamefont {Fujita}}, \bibinfo {author} {\bibfnamefont
  {C.}~\bibnamefont {Reichl}}, \bibinfo {author} {\bibfnamefont
  {W.}~\bibnamefont {Wegscheider}},\ and\ \bibinfo {author} {\bibfnamefont
  {L.~M.}\ \bibnamefont {Vandersypen}},\ }\bibfield  {title} {\bibinfo {title}
  {Loading a quantum-dot based “qubyte” register},\ }\bibfield  {journal}
  {\bibinfo  {journal} {npj Quantum Information}\ }\textbf {\bibinfo {volume}
  {5}},\ \href {https://doi.org/10.1038/s41534-019-0146-y}
  {10.1038/s41534-019-0146-y} (\bibinfo {year} {2019})\BibitemShut {NoStop}%
\bibitem [{\citenamefont {Mills}\ \emph {et~al.}(2019)\citenamefont {Mills},
  \citenamefont {Feldman}, \citenamefont {Monical}, \citenamefont {Lewis},
  \citenamefont {Larson}, \citenamefont {Mounce},\ and\ \citenamefont
  {Petta}}]{Mills2019}%
  \BibitemOpen
  \bibfield  {author} {\bibinfo {author} {\bibfnamefont {A.~R.}\ \bibnamefont
  {Mills}}, \bibinfo {author} {\bibfnamefont {M.~M.}\ \bibnamefont {Feldman}},
  \bibinfo {author} {\bibfnamefont {C.}~\bibnamefont {Monical}}, \bibinfo
  {author} {\bibfnamefont {P.~J.}\ \bibnamefont {Lewis}}, \bibinfo {author}
  {\bibfnamefont {K.~W.}\ \bibnamefont {Larson}}, \bibinfo {author}
  {\bibfnamefont {A.~M.}\ \bibnamefont {Mounce}},\ and\ \bibinfo {author}
  {\bibfnamefont {J.~R.}\ \bibnamefont {Petta}},\ }\bibfield  {title} {\bibinfo
  {title} {Computer-automated tuning procedures for semiconductor quantum dot
  arrays},\ }\bibfield  {journal} {\bibinfo  {journal} {Applied Physics
  Letters}\ }\textbf {\bibinfo {volume} {115}},\ \href
  {https://doi.org/10.1063/1.5121444} {10.1063/1.5121444} (\bibinfo {year}
  {2019})\BibitemShut {NoStop}%
\end{thebibliography}%

\end{document}